\documentclass[%
 aps,
 prl,
twocolumn,
superscriptaddress,
frontmatterverbose, 
showpacs,preprintnumbers,
nofootinbib,
 amsmath,
 amssymb,
floatfix,
latexsym,array,enumerate,letter,
]{revtex4-1}

\usepackage[utf8]{inputenc}
\usepackage[english]{babel}
\pdfoutput=1
\allowdisplaybreaks
\usepackage{graphicx,color}
\usepackage[colorlinks=true,citecolor=blue,linkcolor=blue,urlcolor=blue]{hyperref}
\usepackage{comment}
\usepackage{appendix}
\usepackage{caption}
\usepackage{footmisc}
\usepackage{bm}
\usepackage{dcolumn}
\usepackage{gensymb}
 \usepackage{float} 
\restylefloat{table}
\usepackage{multirow}
 \usepackage{hyperref}
 \usepackage{array}

\hypersetup{
    colorlinks,
    citecolor=black,
    filecolor=black,
    linkcolor=blue,
    urlcolor=blue
}

\newcounter{infer}[section]

\newcolumntype{P}[1]{>{\centering\arraybackslash}p{#1}}
 \newcommand{\be}{\begin{equation}}
\newcommand{\ee}{\end{equation}}
\newcommand{\beq}{\begin{equation}}
\newcommand{\eeq}{\end{equation}}
\newcommand{\bea}{\begin{eqnarray}}
\newcommand{\eea}{\end{eqnarray}}

\usepackage{graphicx,microtype,physics,framed,xcolor}
\usepackage{mdframed,adjustbox} 
\newmdenv[backgroundcolor=gray!15,%
skipabove=0pt,%
skipbelow=5pt,%
leftmargin=0pt,%
rightmargin=0pt,%
innertopmargin=-5pt,%
innerbottommargin=7pt,%
innerleftmargin=2pt,%
innerrightmargin=2pt,%
splittopskip=0pt,%
splitbottomskip=0pt,%
linewidth=0pt,%
nobreak=true]%
{keyeqn}

\begin{document}
\title{Hybrid Strangeon Stars}

\author{Chen Zhang}
\email{iasczhang@ust.hk}
\affiliation{The HKUST Jockey Club Institute for Advanced Study, The Hong Kong University of Science and Technology, Hong Kong S.A.R., P.R.China}

\author{Yong Gao}\email{gaoyong.physics@pku.edu.cn}
\affiliation{Max-Planck-Institut f{\"u}r Gravitationsphysik (Albert-Einstein-Institut),
Am M{\"u}hlenberg 1, D-14476 Potsdam-Golm, Germany}
\affiliation{Department of Astronomy, School of Physics, Peking University, Beijing 100871, China}
\affiliation{Kavli Institute for Astronomy and Astrophysics, Peking University, Beijing 100871, China}

\author{Cheng-Jun Xia}
\email{cjxia@yzu.edu.cn}
\affiliation{Center for Gravitation and Cosmology, College of Physical Science and Technology, Yangzhou University, Yangzhou 225009, China}

\author{Renxin Xu}
\email{r.x.xu@pku.edu.cn}
\affiliation{Department of Astronomy, School of Physics, Peking University, Beijing 100871, China}
\affiliation{Kavli Institute for Astronomy and Astrophysics, Peking University, Beijing 100871, China}

\begin{abstract}
It was conjectured that the basic units of the ground state of bulk strong matter may be strange-clusters called strangeons, and they can form self-bound strangeon stars that are highly compact. Strangeon stars can develop a strange quark matter (SQM) core at high densities, particularly in the color-flavor-locking phase, yielding a branch of hybrid strangeon stars. We explore the stellar structure and astrophysical implications of hybrid strangeon stars. We find that hybrid strangeon stars can meet various astrophysical constraints on pulsar masses, radii, and tidal deformabilities.  Finally, we show that the strangeon-SQM mixed phase is not preferred if the charge-neutrality condition is imposed at the strangeon-SQM transition region.

\end{abstract}
\maketitle
\section{Introduction}

The detection of gravitational waves (GWs) from the coalescence of compact binaries by LIGO/Virgo collaborations~\cite{LIGOScientific:2016aoc,  LIGOScientific:2017bnn,LIGOScientific:2018mvr,TheLIGOScientific:2017qsa,Abbott:2018wiz,Abbott:2020uma,Abbott:2020khf} has greatly improved our knowledge of black holes and compact stars. They offer unique opportunities to probe unconventional QCD matter phases, such as quark matter and strangeon matter.

Quark matter (QM), a state comprised of deconfined free-flowing quarks, can possibly exist inside the neutron star core (i.e. conventional hybrid stars~\cite{Alford:2004pf,Alford:2013aca}). If they are stable at zero pressure, either in form of strange quark matter  (SQM)~\cite{Bodmer:1971we,Witten,Terazawa:1979hq,Farhi:1984qu} or up-down quark 
 matter ($ud$QM)~\cite{Holdom:2017gdc}, they can constitute an entire quark star~\cite{Haensel:1986qb,Alcock:1986hz, Xu:1999bw,Yang:2023haz,Zhang:2019mqb,Zhao:2019xqy, Ren:2020tll,Cao:2020zxi,Wang:2021byk,Yuan:2022dxb,Restrepo:2022wqn,Xia:2020byy,Xia:2022tvx} or
 the crust (i.e. inverted hybrid stars~\cite{Ren:2022loa}), both with potentially distinct astrophysical implications~\cite{Xu:1998ur,Xu:2000sr,Ren-Xin:2002cvr,Wu:2006de,Kankiewicz:2016dha,Kuerban:2019prr,Usov:1997ff,Glendenning:1994zb,Geng:2015uja,Zou:2022voz,Geng:2021apl,Wang:2021jyo,Zhang:2018zdn,Huang:2017ocl,Zhang:2023zth}. Effects from QCD interactions such as color-superconductivity and perturbative QCD (pQCD) corrections can help  quark stars meet various astrophysical constraints~\cite{Weissenborn:2011qu,Zhou:2017pha,Zhang:2020jmb}. Generally, it is expected that at very high densities quark matter should be in the color-flavor-locking phase (CFL), where $u,d,s$ quarks form cooper pairs antisymmetrically in color-flavor space with equal fractions by the attractive one-gluon exchange channel, providing a lowered energy state.


Strangeon matter (SM) is similar to strange quark matter where both are composed of a nearly equal number of $u,d,s$ quarks~\cite{Xu:2003xe, Lai:2009cn, Lai:2017ney,Miao:2020cqj}.
However, strangeon matter has quarks localized as clusters in a globally solid state due to the large masses of and the strong coupling between strangeons. Strangeon stars~\cite{Xu:2003xe,Zhou:2004ue,Zhou:2014tba,Lai:2009cn,Lai:2017ney,Miao:2020cqj,Lai:2017mjv,Lai:2018ugk,Lai:2020mlu,Gao:2021uus} composed of strangeon matter have intrinsic stiff equation of state (EOS) and large compactness, and they had already been proposed to support massive pulsars ($\gtrsim2M_\odot$~\cite{Lai:2009cn}) before the announcement of the first massive pulsar PSR J1614-2230~\cite{Demorest:2010bx}. Recently, we have shown that all strangeon stars are compact enough to feature a photonsphere that is essential to the generation of GW echoes~\cite{Zhang:2023mzb}.

 The transition from strangeon matter to strange quark matter is likely to occur, considering such ``deconfinement" originates from a shrinking of strangeon lattice spacing as density or pressure increases so that the lattice constant becomes smaller than the radius of individual quark bags, as described by the linked bag model in Ref.~\cite{Miao:2020cqj}. This gives rise to a new type of stellar objects, the \emph{Hybrid Strangeon Stars}, consisting of a strangeon crust and a strange quark matter core.
Pure strangeon stars can form from neutron stars absorbing strangeon nuggets, or quantum nucleation in the interior. If SQM is more
stable than SM at some density, then the same process can take place and form hybrid strangeon stars directly or through the SQM quantum nucleation 
inside strangeon stars. Such first-order phase transition needs the center pressure to be larger than some critical value at the corresponding central chemical potential. Such lift of center pressure beyond critical point can happen from spin-down, accretion or merger of strangeon stars.

As for the organization of this paper, we first introduce the 
EOSs of SM and SQM, and constrain the EOS parameters from the stability considerations. Then, with Maxwell constructions where a sharp interface is assumed, we solve the hybrid stellar structures and study their compatibility with astrophysical constraints. Finally, we explore the possibility of mixed phase (Gibbs construction) where the transition region is with mixed SM and SQM rather than a sharp interface.
 


\section{Equations of States}
For the quark matter sector, we adopt the unified treatment of interacting quark matter that recently developed in~\cite{Zhang:2020jmb} and later  applied in several studies~\cite{Zhang:2021fla,Zhang:2021iah,Blaschke:2021poc,Blaschke:2022egm,Koliogiannis:2022uim,Oikonomou:2023otn,Gammon:2023uss}.

Referring to~\cite{Zhang:2020jmb}, we first rewrite the thermodynamic potential $\Omega$ of the superconducting quark matter~\cite{Alford:1998mk,Rajagopal:2000ff,Lugones:2002va,Alford:2002kj,Alford:2004pf} in a general form with the pQCD correction~\cite{Fraga:2001id} included:  
\begin{equation}\begin{aligned}
\Omega=&-\frac{\xi_4}{4\pi^2}\mu^4+\frac{\xi_4(1-a_4)}{4\pi^2}\mu^4- \frac{ \xi_{2a} \Delta^2-\xi_{2b} m_s^2}{\pi^2}  \mu^2  \\
&-\frac{\mu_{e}^4}{12 \pi^2}+B ,
\label{omega_mu}
\end{aligned}\end{equation}
where $\mu$ and $\mu_e$ are the respective average quark and electron chemical potentials. The first term represents the unpaired free quark gas contribution. The second term with $(1-a_4)$ represents the pQCD contribution from one-gluon exchange for gluon interaction to $O(\alpha_s^2)$ order. To phenomenologically account for  higher-order contributions, we can vary $a_4$ from $a_4=1$, corresponding to a vanishing pQCD correction, to very small values where these corrections become large~\cite{Fraga:2001id,Alford:2004pf,Weissenborn:2011qu}. The term with $m_s$ accounts for the correction from the finite strange quark mass if applicable, where $m_s=95\pm 5 \rm \, MeV$~\cite{ParticleDataGroup:2014cgo}, and we choose $m_s=95 \rm \, MeV$ as its benchmark value.
The term with the gap parameter $\Delta$ represents the contribution from color superconductivity.  $(\xi_4,\xi_{2a}, \xi_{2b})$ represents different state of color-superconducting phases.  $B$ is the effective bag constant that accounts for the nonperturbative contribution from the QCD vacuum.


The corresponding equation of state was derived in Ref.~\cite{Zhang:2020jmb}:
\be
P=\frac{1}{3}(\rho-4B)+ \frac{4\lambda^2}{9\pi^2}\left(-1+\text{sgn}(\lambda)\sqrt{1+3\pi^2 \frac{(\rho-B)}{\lambda^2}}\right),
\label{eos_tot}
\ee
where \be
\lambda=\frac{\xi_{2a} \Delta^2-\xi_{2b} m_s^2}{\sqrt{\xi_4 a_4}}.
\label{lam}
\ee 
 Note that $\rm sgn(\lambda)$ represents the sign of $\lambda$. 
 The chemical potential (per baryon number) has the  following form:
\be
\mu_{\rm QM}=\frac{3\sqrt{2}}{(a_4 \xi_4)^{1/4}}\sqrt{[(P+B)\pi^2+\lambda^2]^{1/2}-\lambda}\,.
\label{muQ}
\ee
 Taking the zero pressure limit of $\mu_{\rm QM}$, we obtain the energy per baryon number, which can be converted into  the following form:
\be
\left(\frac{E}{A}\right)_{\rm QM}=\frac{3\sqrt{2} \pi}{(\xi_4 a_4)^{1/4}}\frac{ {B}^{1/4}}{\sqrt{(\lambda^2/B+\pi^2)^{1/2}+\lambda/ \sqrt{B}}}, 
\label{EA_Q}
\ee
where we see a larger $\lambda$ lowers the energy as  expected.

  We have examined that hybrid strangeon star with a core of unpaired strange quark matter ($\Delta=0$) cannot support $2 M_{\odot}$ while retaining radial stability that requires $\partial M/\partial P_{c} >0$.
  This is not a surprise considering strangeon EOS is much stiffer than that of unpaired SQM, and a transition to a much softer EOS is likely to induce radial instabilities due to insufficient degenerate pressure to resist the gravitational pulling. We can thus stabilize the hybrid strangeon stars by introducing color-superconductivity effects to stiffen the SQM EOS~\footnote{Such instabilities can also be cured by considering the scenario of slow SM-SQM conversions (with respect to radial-oscillation timescale)~\cite{Pereira:2017rmp,Lugones:2021zsg}. Considering both SM and CFL have the three-flavor symmetry, we expect that the surface tension of SM-SQM is not large, thus the conversion is likely to be fast and correspondingly the stability criteria retains to be $\partial M/\partial P_c>0$, i.e. the star mass increases with center pressure.}. Therefore, in the following discussions, we specify the SQM phase to be CFL $(\xi_4=3,\xi_{\rm 2a}=3, \xi_{\rm 2b}=3/4)$, considering the shared flavor composition and the fact that color superconductivity stiffens the EOSs.  Besides, we set $a_4=1$ (no extra QCD corrections) for simplicity.

Following previous studies~\cite{Xu:2003xe,Zhou:2004ue,Zhou:2014tba,Lai:2009cn,Lai:2017ney,Miao:2020cqj,Lai:2017mjv,Lai:2018ugk,Lai:2020mlu,Gao:2021uus}, we assume the interaction potential between two strangeons is described by the Lennard-Jones potential~\cite{Jones(1924)}:
\be
U(r)= 4\epsilon \left[ \left(\frac{\sigma}{r}\right)^{12}-\left(\frac{\sigma}{r}\right)^6 \right],
\ee
where $r$ is the distance between two strangeons, and $\sigma$ is the distance when $U(r)=0$. The parameter $\epsilon$ describes the depth of the interaction potential between strangeons. A larger $\epsilon$ will then indicate a larger repulsive force at short range and thus maps to a stiffer EOS.

The mass density $\rho$ and pressure $p$ of zero-temperature dense matter composed of strangeons derived from the Lennard-Jones potential~\cite{Lai:2009cn} reads
\bea\label{eq:energy_density}
    \rho&=&2 \epsilon\left(A_{12} \sigma^{12} n^{5}-A_{6} \sigma^{6} n^{3}\right)
    + nN_{\rm q}m_{\rm q}\,, \\
     P&=&n^{2} \frac{\mathrm{d}(\rho / n)}{\mathrm{d} n}=4 \epsilon\left(2 A_{12}
    \sigma^{12} n^{5}-A_{6} \sigma^{6} n^{3}\right)\,,
\eea
where $A_{12}=6.2$, $A_{6}=8.4$, and $n$ is the number density of strangeons. $N_{\rm q}m_{\rm q}$ is the mass of a strangeon with $N_{\rm q}$ being the
number of quarks in a strangeon and $m_{q}$ being the average constituent quark mass. The contributions from degenerate electrons and vibrations of the
lattice are neglected due to their expected smallness.

At the surface of  strangeon stars, the pressure becomes zero, and we obtain the surface
number density of strangeons as $\big[A_{6}/(2A_{12}\sigma^{6})\big]^{1/2}$. For
convenience, it is transformed into baryon number density, i.e.,
\begin{equation} \label{eq:surface}
    n_{\rm s}=\left(\frac{A_{6}}{2A_{12}}\right)^{1/2}\frac{N_{\rm
    q}}{3\sigma^{3}}\,,
\end{equation}
so that the EOS can be rewritten into the following simpler form
\be
\begin{aligned}
\frac{\rho}{ n_s}&= \frac{a}{9}  \tilde{\epsilon}  \left(\frac{1}{18  } \bar{n}^5 - \bar{n}^3\right)+ m_q \bar{n}, \\
\frac{P}{ n_s}&=\frac{2 \,a}{9}  \tilde{\epsilon}  \left(\frac{1}{9  }\bar{n}^5 -\bar{n}^3\right),
\end{aligned}
\label{EOS_strangeon}
\ee
where $a=A_6^2/A_{12}=8.4^2/6.2\approx11.38$,  $\tilde{\epsilon}=\epsilon/N_q$ and $\bar{n}=N_q \,n / n_s$. Note that $\bar{n}=3$ at star surface where $P=0$.

The chemical potential of strangeon matter can be derived via the thermodynamic relation $\mu=(\rho+P)/n$. Note that to study its crossings with $\mu_{\rm QM}$, one needs to further convert it to the chemical potential per baryon number 
\be
\mu_{\rm strangeon}=\frac{3\mu}{N_q}=3\frac{\rho/n_s+P/n_s}{\bar{n}}=3m_q+{a}\tilde{\epsilon}(\frac{5}{54}\bar{n}^4-\bar{n}^2).
\label{muS}
\ee
 Referring to Eq.~(\ref{EOS_strangeon}), we see that both the EOS $P(\rho)$ and $\mu_B(P)$ for strangeons only depends on parameters $n_s$ and  $\tilde{\epsilon}$ with the dependence on $N_q$ absorbed. Taking the zero pressure limit of $\mu_{\rm strangeon}$, we obtain the energy per baryon number at the bulk limit: 
\be
\left(\frac{E}{A}\right)_{\rm strangeon}=3m_q-\frac{3a}{2}\tilde{\epsilon},
\label{EA_S}
\ee
where we see that $E/A$ has no dependence on $n_s$, decreases as $\tilde{\epsilon}$ increases,  so that strangeon matter can be the ground state of matter at the bulk limit for a finite $\tilde{\epsilon}$. In this proof-of-concept work, we adopt $3m_q=930$ MeV for simplicity, ensuring that, at the bulk limit, strangeon matter is always more stable than nucleon matter, since $(E/A)_{\text{strangeon}}<(E/A)_{\rm Fe}=930\rm \,MeV$, the energy per baryon number of the most stable nucleus ${}^{56}\text{Fe}$.
Requiring a positive $E/A$ (or a non-negative $\rho$ at zero pressure) sets a theoretical bound: $\epsilon/N_q\leq 2m_q/a\approx 54.5 \rm \,MeV.$

\begin{figure}[h]
 \centering
\includegraphics[width=8cm]{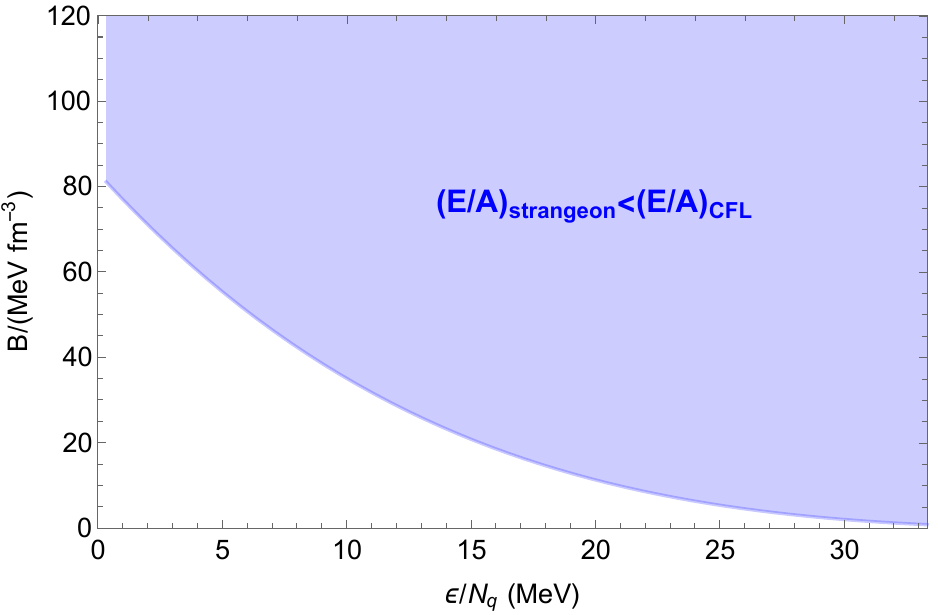}  
\includegraphics[width=8cm]{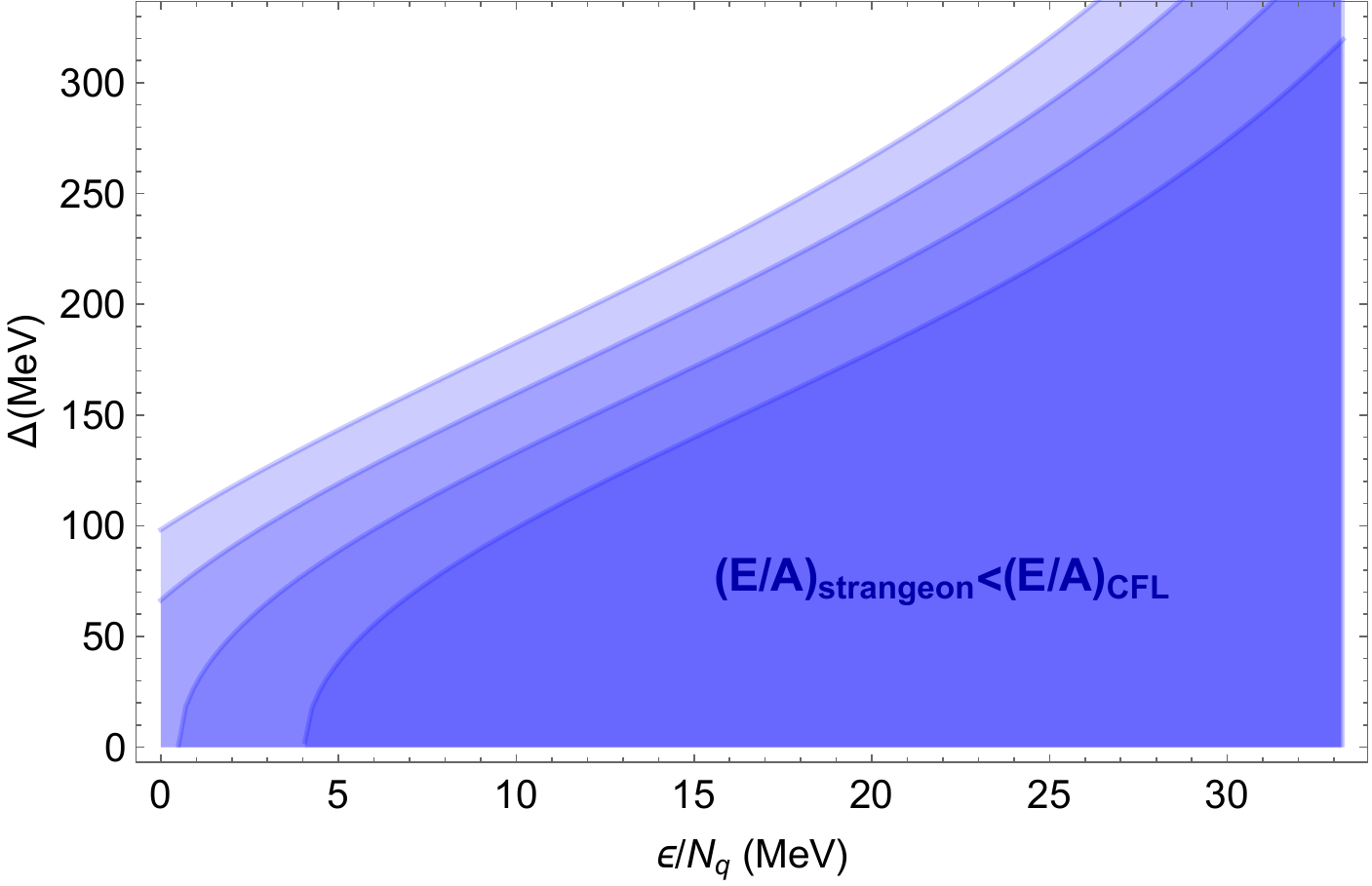}  
\caption{Allowed parameter space (blue-shaded) for the existence of {hybrid strangeon stars} from stability consideration  $(E/A)_{\rm Strangeon}<(E/A)_{\rm CFL}$. Top: CFL bag constant $B$ and bottom: CFL superconductivity gap $\Delta$ versus parameter $\epsilon/N_q$ of strangeon matter. For the bottom sub-figure, the shaded region with lighter-colored contour lines represents larger bag constant, sampling $B=60, 80, 100, 120\, \rm MeV/fm^3$ (bottom to top).  }
   \label{paraHyb}
\end{figure}

The transition pressure or density can be determined by the crossings of their chemical potentials. A necessary condition for such chemical potential crossing is that the zero-pressure chemical potential (i.e., energy per baryon number $E/A$) of strangeon (Eq.~(\ref{EA_S})) is smaller than that of CFL (Eq.~(\ref{EA_Q})). 
We show the related parameter space as the blue-shaded bands of Fig.~\ref{paraHyb}. We see that overall, the existence of such a hybrid configuration prefers a relatively stiff strangeon EOS (large $\tilde{\epsilon}$) but a relatively soft CFL phase (large $B$ or small $\Delta$). 

On the other hand, the hybrid configuration would become radially unstable ($\partial M/\partial P_{\rm c}<0$) if the transition pressure $P_{\rm trans}$ is too large~\cite{Alford:2013aca},  as we have also examined explicitly. Referring to Eq.~(\ref{muQ}) and Eq.~(\ref{muS}), we see that the strangeon matter to SQM transition is more likely to occur at smaller $P_{\rm trans}$ in the case of a smaller $B$, a larger $\Delta$ (stiffer SQM EOS), a smaller $\tilde{\epsilon}$ (softer SM EOS) or smaller $n_s$. These conditions compete with those from the stability condition mentioned in the previous paragraph, constraining the allowed parameter space.

\begin{figure*}[htb]
 \centering
\includegraphics[width=8  cm]{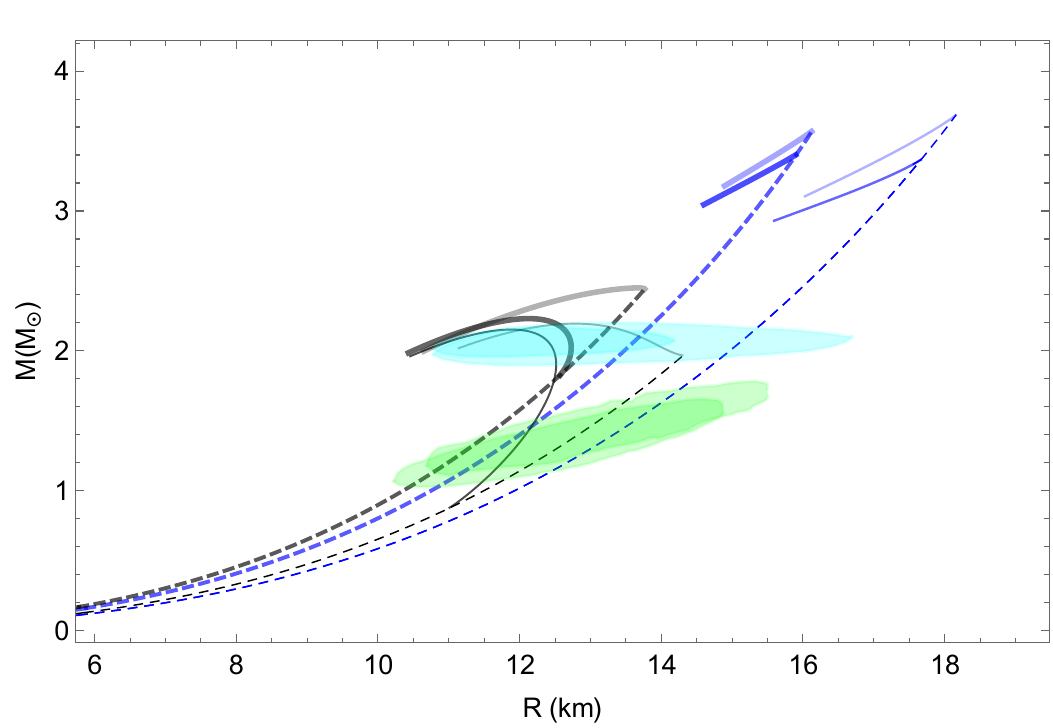}  
\includegraphics[width=8 cm]{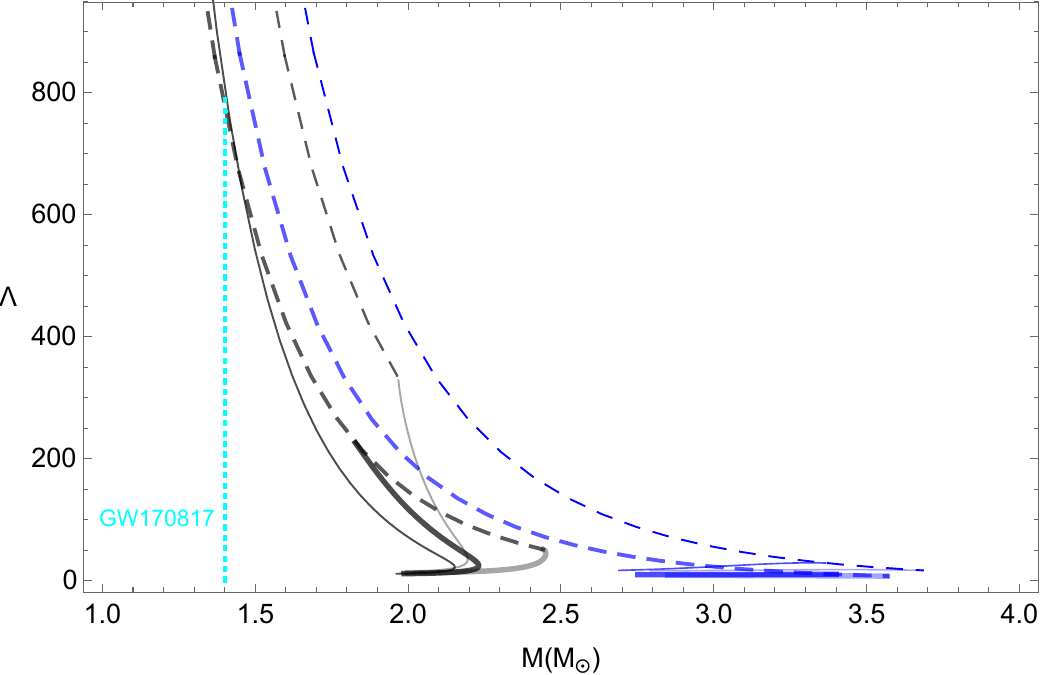}  
\includegraphics[width=8  cm]{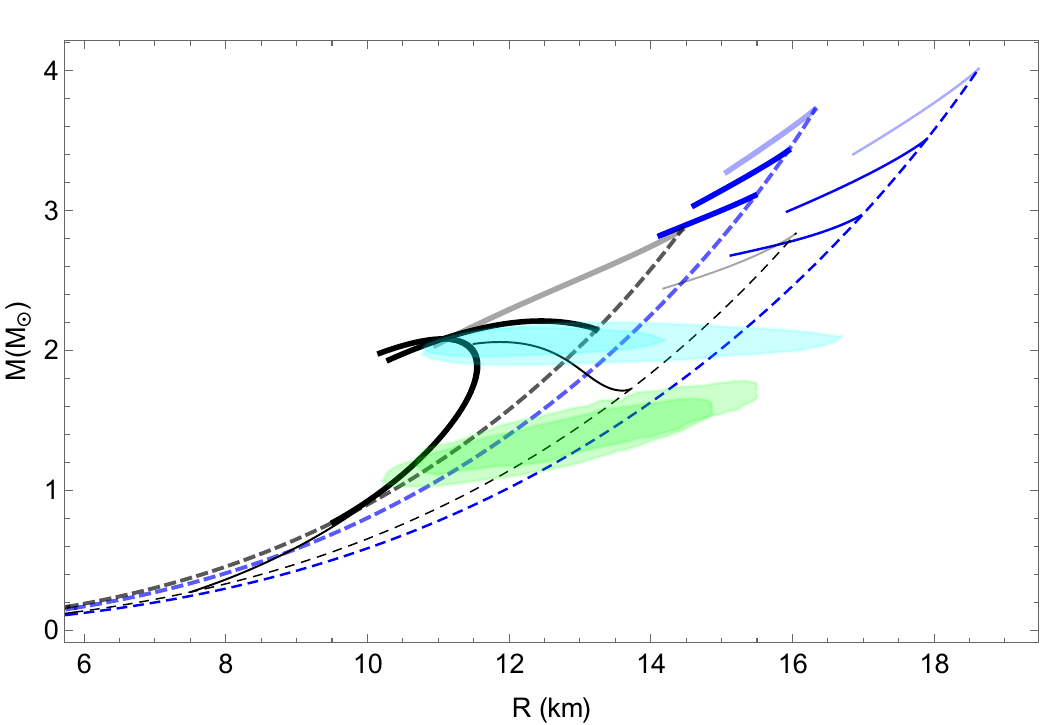}  
\includegraphics[width=8 cm]{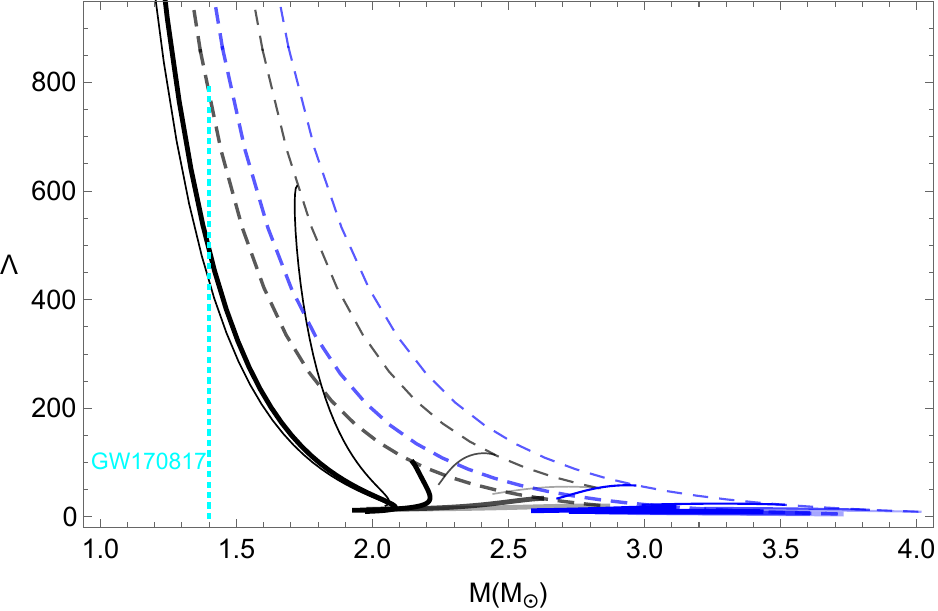}  
\caption{The curves $M$-$R$ (left) and $\Lambda$-$M$ (right) of hybrid strangeon stars (solid lines) with $\epsilon/N_q=80/9\approx 8.9$  (black), $120/9\approx13.3$ (blue) MeV, $n_s=0.22$ (thin), $0.30$ (thick) $\rm fm^{-3}$ for the strangeon composition, and $B=60$ (top), $80$ (bottom) $\rm \, MeV/fm^3$ for the CFL composition. Lines with darker colors denote larger $\Delta$, sampling  60, 80 MeV for top panels and 60, 100, 120 MeV for bottom panels, respectively. (no large-$\Delta$ lines in top panels due to stability constraints referring to Fig.~\ref{paraHyb}.) Dashed lines are pure strangeon star configurations. Shaded regions are constraints with $90\%$ credibility from the NICER mission PSR J0030+0451 (green colored) \cite{NICER1,NICER2}, PSR J0740+6620 (cyan colored)\cite{NICER3,NICER4}. The cyan-dotted vertical line in the right panels denotes the GW170817's $\Lambda(1.4 M_{\odot})\leq 800$ constraint~\cite{TheLIGOScientific:2017qsa}. }
   \label{MRHyb}
\end{figure*}
\section{astrophysical implications} 
The stellar structure can be solved via the Tolman-Oppenheimer-Volkoff (TOV) equation~\cite{Tolman:1939jz,Oppenheimer:1939ne},
 \bea
 \begin{aligned}
\frac{d m}{d r}& = 4 \pi  \rho r^2\,,\label{eq:dm}\\
\frac{d P}{d r} &= (\rho+P)  \frac{m + 4 \pi P r^3}{2 m r -r^2},\,\\
\end{aligned}
\label{tov}
\eea
where the profiles $P(r)$ and $m(r)$ are solved as functions of the center pressure $P_{\rm c}$. The radius $R$ and physical mass $M$ of the compact stars are determined by $P(R)=0$ and $M=m(R)$, respectively. One then obtains the mass-radius relation $M(R)$ of hybrid strangeon stars by solving the TOV equations together with the EOSs of the two matter phases, where the transition point is determined by the crossing of their chemical potentials, as introduced in the last section.

To compare with gravitational wave observations, we can further compute the dimensionless tidal deformability $\Lambda=2k_2/(3C^5)$, where  $C=M/R$ is the compactness and $k_2$ is the Love number that characterizes the stars' response to external disturbances~\cite{AELove,Hinderer:2007mb,Hinderer:2009ca,Postnikov:2010yn}.  
The Love number $k_2$ can be determined by solving a function $y(r)$ from a specific differential equation \cite{Postnikov:2010yn} and the TOV equation Eq.~(\ref{tov}), with the boundary condition $y(0)=2$. For hybrid configurations, the matching condition~\cite{Damour:2009vw,Takatsy:2020bnx} $y(r_{d}^+) - y(r_{d}^-) = -4\pi r_{d}^3 \Delta \rho_d /(m(r_{d})+4\pi r_{d}^3\, P(r_{d}))$ should be imposed at $r_d$ (i.e., the core radius and the star radius), where an energy density jump $\Delta \rho_d$ occurs. 

 For illustration, we show various benchmark TOV solutions and corresponding tidal deformabilities in Fig.~\ref{MRHyb} for $B=60, 80$ $\rm MeV/fm^3$ with $\tilde{\epsilon}$ and $\Delta$ choices satisfying  $(E/A)_{\rm Strangeon}<(E/A)_{\rm CFL}$  (shaded bands in Fig.~\ref{paraHyb}). 
 
 We see that all the benchmark examples shown in Fig.~\ref{MRHyb} satisfy NICER constraints \footnote{Note that we show the NICER X-ray constraints in the graph but neglect them in the table considering the X-ray analyses of hybrid strangeon stars may be different from those of neutron stars.}, while the GW170817 constraints ($\Lambda_{1.4 M_{\odot}}\leq800$) can be met by hybrid strangeon stars with $(\tilde{\epsilon},n_s\rm /fm^{-3}, \Delta/MeV)=(80/9,0.22, 80)$ for $B=60\rm \, MeV/fm^3$ (upper panels), and $ (\tilde{\epsilon},n_s\rm/fm^{-3}, \Delta/MeV)=(80/9,0.22, 120)$, $(80/9,0.3, 120)$ for $B=80\rm \, MeV/fm^3$ (lower panels).

 The general features of correlations between constraints and parameters are summarized in Table~\ref{tab:trend}. For example, as the second row of Table~\ref{tab:trend} summarizes, hybrid strangeon stars with small  $\tilde{\epsilon}$ (black lines) and $n_s$ (thin lines), or large $\Delta$ (darker colored lines) and small B (such as upper panels) tend to be radially unstable ($\partial M/\partial P_{\rm c}<0$), which means radial stabilities require CFL to be not too soft compared to the stiffness of strangeon EOS, considering $\Delta$ and $\epsilon/N_q$ signal the stiffness of each of the two matter phases.
However, we also see that the $M_{\rm TOV}\gtrsim 2 M_{\odot}$ constraint~\cite{Demorest:2010bx} prefers overall stiff EOSs for both two matter phases (a large $\tilde{\epsilon}$ or $\Delta$), while GW170817 tidal deformability constraint ($\Lambda_{1.4 M_{\odot}}\leq800$~\cite{TheLIGOScientific:2017qsa}) prefers the opposite at low center densities. These together set bounds on the allowed parameter space. 

\begin{table}[htbp]
\begin{center}
\small
\resizebox{\columnwidth}{!}{%
\begin{tabular}{cccccc}
\hline\hline
     & $\tilde{\epsilon}$
     & $n_s$ 
     &  $\Delta$
     &  $B$ \\
\hline
\\
&&&&&
\\[-6.5mm]
 $(E/A)_{\rm Strangeon}<(E/A)_{\rm CFL}$\,\, & \,\,$+$\,\,  & \,\,$\symbol{92}$ \,\, &  \,\, $-$\,\,  & \,\, $+$\,\,  \\ 
 \hline
\\
&&&&&
\\[-6.5mm]
$\partial M/\partial P_{\rm c}>0$\,\, & \,\,$-$\,\,  & \,\,$-$ \,\, &  \,\,$+$ \,\,  & \,\, $-$\,\,  \\ 
 \hline
\\
&&&&&
\\[-6.5mm]
 $M_{\rm TOV}\gtrsim 2 M_{\odot}$  \,\, & \,\, $+$ \,\,  & \,\,$-$\,\, & \,\,$+$ \,\,  & \,\, $-$\,\,  \\  
\hline
\\
&&&&&
\\[-6.5mm]
 $\Lambda_{1.4 M_{\odot}}\leq 800$\,\, & \,\,$-$\,\,  & \,\,$+$ \,\, &  \,\,$-$\,\,  & \,\, $+$\,\,  &   \\
 
\hline\hline 

\end{tabular}
}
\caption{Correlations of constraints and the EOS parameters. Plus(minus) sign means positive  (negative) correlation, while slash sign  means no correlation. }
\label{tab:trend}
\end{center}  
\end{table}\vspace{-0ex}

\begin{figure}[h]
 \centering
\includegraphics[width=8cm]{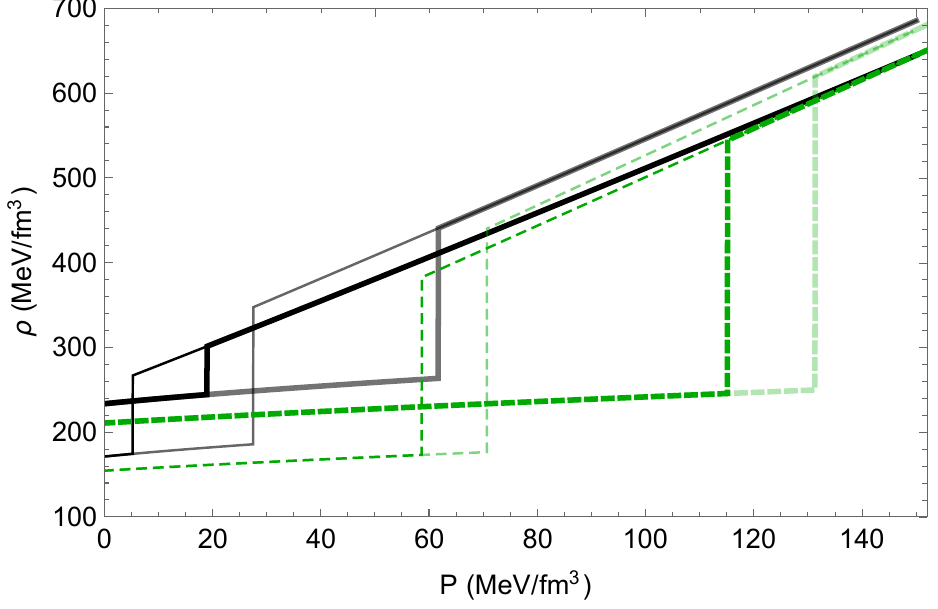}  
\caption{The relations of $\rho(P)$ for stable hybrid strangeon stars, with $\epsilon/N_q=80/9$ MeV, $n_s=0.22$ (thin), $0.30$ (thick) $\rm fm^{-3}$ for the strangeon composition, and $B=60$ (green dashed), $80$ (black solid) $\rm \, MeV/fm^3$ for the CFL composition. Lines with darker colors denote larger $\Delta$, sampling  60, 80 MeV for green lines and 100, 120 MeV for black lines, respectively.}
   \label{rhoP}
\end{figure}

For completeness,  we plot in Fig.~\ref{rhoP} the EOSs of the branches of Fig.~\ref{MRHyb} that have stable hybrid strangeon stars. We see that the transition mass density $\rho_{\rm trans}$, as determined from chemical potential crossing, is around the nuclear saturation mass density ($\rho_{\rm sat}\approx157\rm\, MeV/fm^3$) for $n_s=0.22\,\rm/fm^3$ lines, and increases to $1.5\rho_{\rm sat}$ for $n_s=0.30\,\rm/fm^3$ lines. The mass density jumps at the transition points vary from $100\,\rm MeV/fm^3$ to $370\,\rm MeV/fm^3$, mainly affected by the variations of parameter $\Delta$.

\begin{figure*}[htb]
 \centering
 \includegraphics[width=8cm]{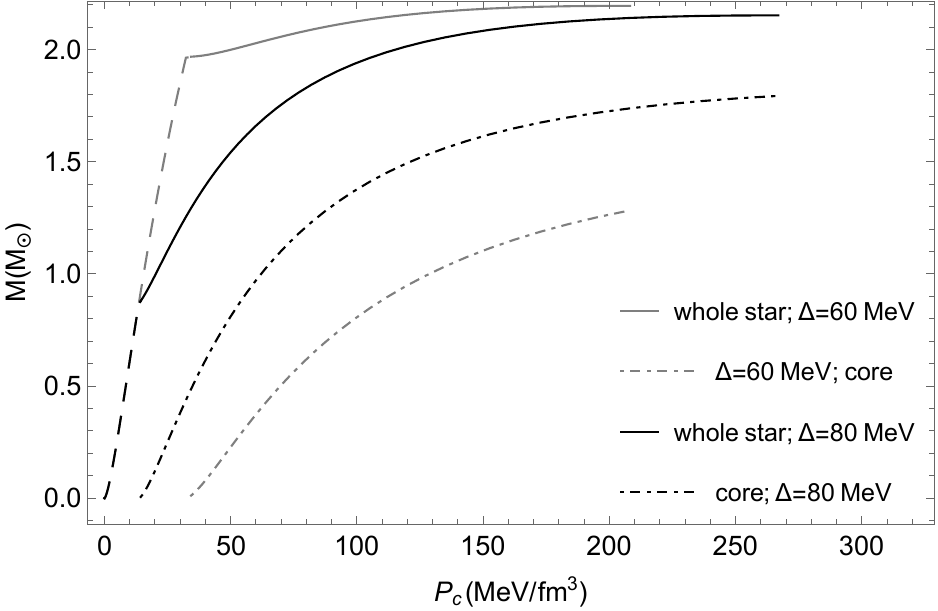}  
\includegraphics[width=8cm]{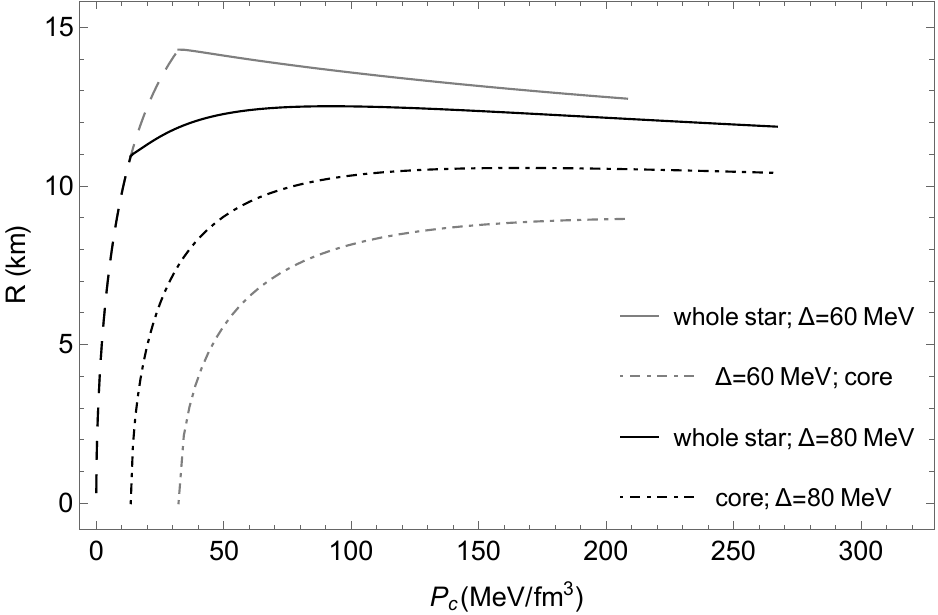}  
\includegraphics[width=8cm]{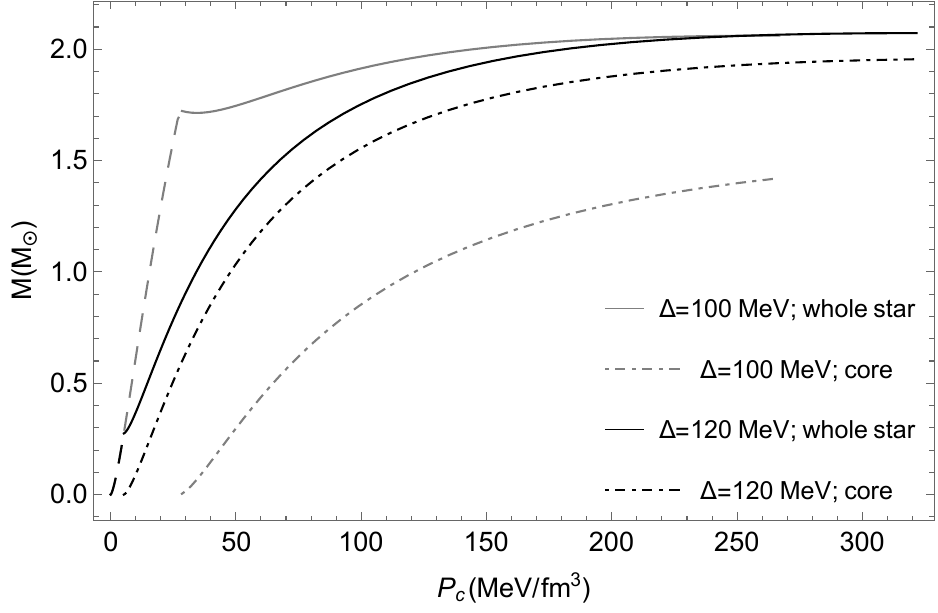} 
\includegraphics[width=8cm]{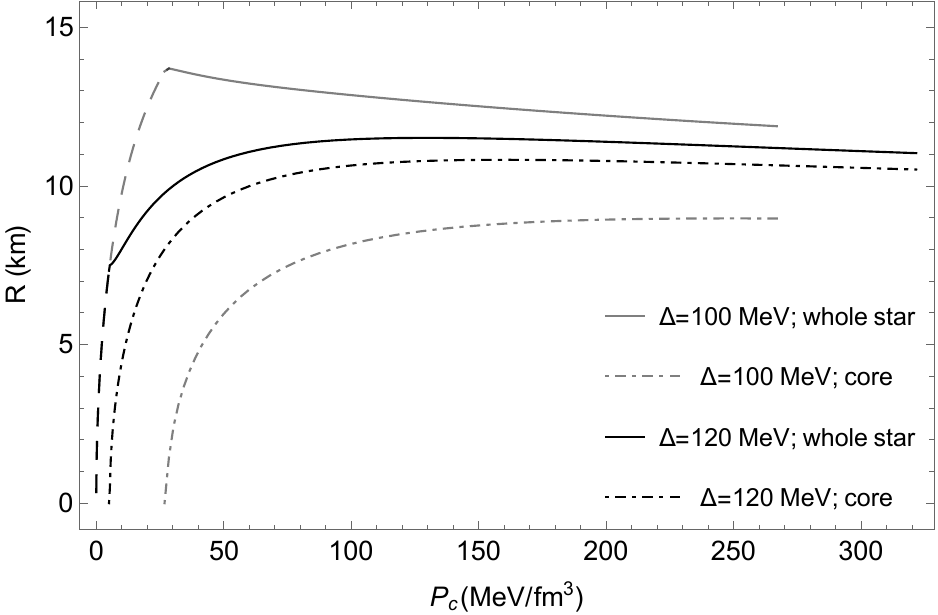} 
\caption{Mass (left) and radius (right) versus center pressure, $P_{\rm c}$, for hybrid strangeon stars with strangeon crusts of \{$n_s=0.22\rm /fm^3, \tilde{\epsilon}=80/9\, MeV$\}, and CFL cores of 
\{$B=60 \,\rm MeV/fm^3$,  $\Delta= 60, 80$ MeV\} (top) and  
\{$B=80 \,\rm MeV/fm^3$,  $\Delta= 100, 120$ MeV\} (bottom).  Darker color denotes larger $\Delta$ values. Dashed lines denote pure strangeon stars. Solid lines denote hybrid strangeon stars. The dot-dashed lines denote the CFL cores. The right ends of the solid and dot-dashed lines are truncated at the corresponding maximum mass points.} 
   \label{fig_MRP}
\end{figure*}

To elaborate on the explicit layer structure, we dissect hybrid strangeon stars by showing the masses and radii of their CFL cores as functions of the centre pressure in Fig.~\ref{fig_MRP}. Here, we choose two benchmark examples of bag constant $B=60, 80 \,\rm MeV/fm^3$ with different CFL gaps and a fixed strangeon phase ($n_s=0.22\rm /fm^3, \tilde{\epsilon}=80/9\, MeV$). We find that, as a general feature, the compact stars are pure strangeon stars at low $P_{c}$, and then develop a CFL core as $P_{c}$ increases. 
At the maximum mass points, the strangeon crusts have widths of $1\sim 5$ km and masses of $ 0.1\sim1\, M_\odot$, where a smaller bag constant or a smaller $\Delta$ maps to a thicker crust.  At the $M=2\, M_{\odot}$ point, all cases map to 
hybrid strangeon stars, with a core of mass $1.13\, (1.86) \, M_\odot$ for $\Delta=100\, (120)$ MeV case when $B=80$ $\rm MeV/fm^3$, and  a core of mass $0.27 \,(1.50) \, M_\odot$ for $\Delta=60\, (80)$ MeV case when $B=60\, \rm MeV/fm^3$. 
At the $M=1.4\, M_{\odot}$ point, for $B=60\, (80)\, \rm MeV/fm^3$,  the $\Delta=60\, (100)$ MeV case is a pure strangeon star, while the $\Delta=80\, (120)$ MeV case is a hybrid strangeon star with a core of mass $0.64\, (1.16) M_{\odot}$.

 
\section{Mixed Phase}
A strangeon-quark mixed phase is possible around the intersurface of the strangeon crust and quark matter core, in analogy to the hadron-quark mixed phase in the conventional hybrid neutron stars.

To construct mixed phase of hybrid strangeon stars, we need $\mu_e\neq0$, thus the strange quark matter sector should be in either the normal unpaired phase ($\Delta=0$)\footnote{ As aforementioned for the sharp transition, normal strange quark matter core with a strangeon crust are not likely to be radially-stable. We expect the situation may be alleviated in the mixed phase scenario.} or the charged CFL phase~\cite{Alford:2004pf}, where $s$ quarks no longer have an equal fraction as $u,d$ quarks. We keep the flavor-symmetry in the strangeon sector considering its solid state with charge-neutrality being enforced, since the Compton wavelength of dilute electrons is much larger than the scale of a strangeon.

\begin{figure}[h]
 \centering
 \includegraphics[width=8cm]{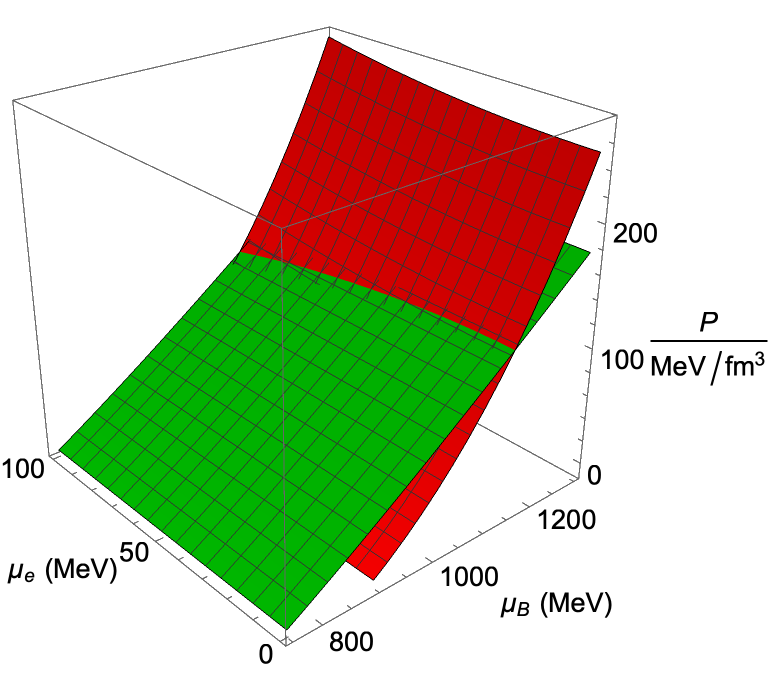}  
 \includegraphics[width=8cm]{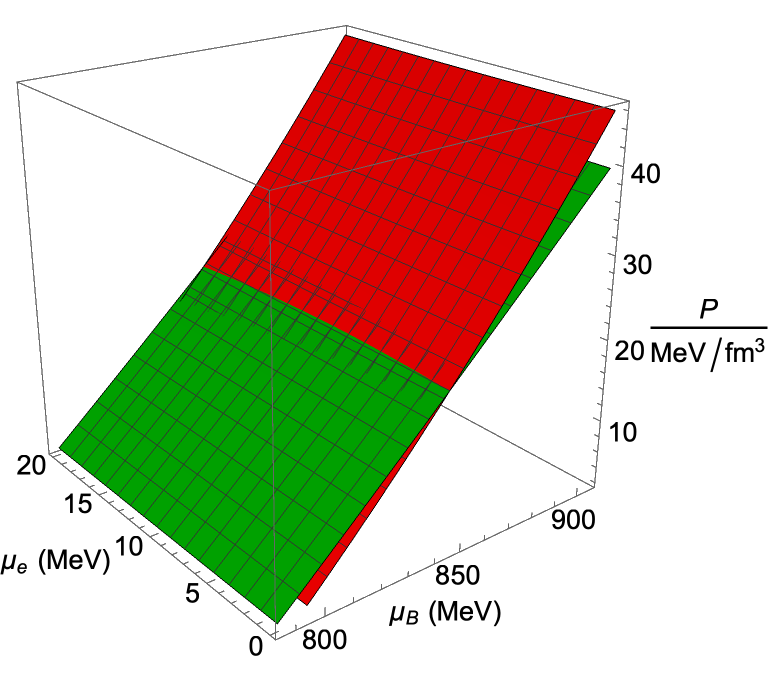}  
\caption{Pressure is plotted as a function of  $\mu_B$ and $\mu_e$ for strangeon phase (green) and strange quark matter (red) of normal unpaired (top panel) and charged CFL phase (bottom panel) of $\Delta=100$ MeV. The mixed phase sits in the intersection of the two surfaces. For illustration, here $B=80\rm \, MeV/fm^3$, $m_s=95 \rm \, MeV$ for the SQM phase and $\epsilon/N_q=80/9, n_s=0.3\rm\,/fm^3$ for strangeon phase. }
   \label{mixed}
\end{figure}

We adopt here the Gibbs construct for the mixed phase as outlined in Ref. \cite{Schertler:1999}. In this case, one may achieve the charge neutrality where the pressures of both strangeon and quark matter are functions of the baryon and electron chemical potentials $\mu_B$ and $\mu_e$. The Gibbs condition for the equilibrium between the two phases (at zero temperature) is
\begin{equation} \label{gibbs.condition}
P_{\rm{SnP}}(\mu_B,\mu_e) = P_{\rm{QkP}}(\mu_B,\mu_e) = P_{\rm{MxP}}(\mu_B,\mu_e),
\end{equation}
where the pressure function for strangeon phase $P_{\rm{SnP}}(\mu_B)$ can be inferred from Eq.~(\ref{EOS_strangeon}) and Eq.~(\ref{muS}) with addition of background electrons $P_{\rm{SnP}}(\mu_B,\mu_e)=P_{\rm{SnP}}(\mu_B)+\mu_{e}^4/(12 \pi^2)$.  Besides, for quark matter phase $P_{\rm{QkP}}$ can be inferred from Eq.~(\ref{omega_mu}) with the identities $p=-\Omega$, $\mu=\mu_B/3$. Their intersection yields the mixed phase $P_{\rm{MxP}}(\mu_B,\mu_e)$, as shown in Fig.~\ref{mixed}.

The global charge neutrality condition reads
\begin{eqnarray} \label{global.charge.neutrality}
 (1-\chi)~n_c^{\rm SnP} +\chi~n_c^{\rm QkP} = 0,
\end{eqnarray}
where, $n_c^{\rm SnP}$ and $n_c^{\rm QkP}$ denote the total charge densities in strangeon phase and quark matter phase (either unpaired SQM or CFL) respectively,  with
\bea
n_c^{\rm SnP}&=&-n_c^{\rm e},\\
n_c^{\rm QkP}&=&\frac{2}{3}n_u-\frac{1}{3}n_d-\frac{1}{3}n_s-n_c^{\rm e},
\eea
where $n_c^{\rm e}=\mu_e^3/(3\pi^2)$, $n_{u,d}=\mu_{u,d}^3/\pi^2$, $n_{s}=(\mu_s^2-m_s^2)^3/\pi^2$ with $\mu_i$ the quark chemical potential of flavor $i$. $\chi$ defines the volume fraction of the quark matter in mixed phase defined as $\chi=V_{\rm QkP}/(V_{\rm QkP}+V_{\rm SnP}).$

We have examined various combinations of parameter sets for SQM in either the normal unpaired phase or CFL phase, finding that the mixed phase $ P_{\rm{MxP}}(\mu_B,\mu_e)$ that satisfies the global charge neutrality condition only resides in a very tiny segment of the intersection line, with variations of $\mu_B$ smaller than $1$ MeV range near the zero $\mu_e$ point, where $\mu_e$ lift to $6\sim8$ MeV. Thus the mixed-phase region for hybrid strangeon stars is negligible, and all results should approximately be the same as those obtained from the Maxwell construction studied in the last section, i.e., the system is effectively reduced into one conserved charge due to the negligible contribution of electrons. As we have examined, introducing QCD corrections ($a_4<1$) will lift the intersecting $\mu_B$ but does not help enlarge the charge-neutral region of the mixed phase. This matches the expectation that the flavor-symmetry breaking effects are small in both strangeon and SQM sectors, resulting in a very small $\mu_e$ and its limited variation range when considering charge neutrality.

\section{Summary and discussion}
We have explicitly shown the new possibility of the hybrid configuration of strangeon stars with a strange quark matter core and a thick strangeon crust. We also demonstrated their compatibility with astrophysical constraints with selected benchmark examples.   It is  shown that mixed phase is not preferred for hybrid strangeon stars with a CFL core.

Hybrid strangeon stars can naturally accommodate the pulsar glitch phenomena as a result of the star quakes in the thick strangeon crust~\cite{Lai:2017xys,Wang:2020xsm,Lai:2023axr,Lu:2023dwi}, in contrast to compact stars with the crystalline color superconducting phase where glitches are a result of superfluid vortices pinned to the solid component~\cite{Anglani:2013gfu,Lin:2013nza,Mannarelli:2014ija,Mannarelli:2015jia}. The large density discontinuity at the SM-CFL intersurface (referring to Fig.~\ref{rhoP}) will induce $g$-mode gravitational waves from nonradial oscillations that might help differentiate hybrid 
 strangeon stars and other types of compact stars~\cite{Flores:2013yqa,Ranea-Sandoval:2018bgu,Constantinou:2021hba,Constantinou:2023ged,Zhao:2022toc}. Besides, the large shear modulus change and density continuities at the crust-core interface are likely to result in large and distinct crust-core interfacial modes that can also be probed by gravitational-wave observations~\cite{Lau:2020bfq, Zhu:2022pja}. 

As aforementioned, the conversions from strangeon matter to strange quark matter (either unpaired or CFL) is likely to be fast from intuitive expectations. However, due to the intrinsic uncertainties of the SM-SQM surface tension in nonperturbative QCD, it is not entirely impossible that the conversion is slow compared to the radial oscillation  timescale,  corresponding to the slow conversion scenario where branches of $\partial M/\partial P_c<0$ can also be stable. In  this  case, referring to Table~\ref{tab:trend}, the allowed parameter space for stable hybrid strangeon stars can be enlarged. In particular, superconducting gap $\Delta$ can now have smaller values without ruining the radial stability. Similar relaxation of parameter constraints would also be possible when considering the merger remnants with extremely spinning where the unstable branches may have gravitational-wave signals. We leave these for future studies.


\begin{acknowledgments}
\noindent\textbf{Acknowledgments. }  
 C. Zhang is supported by the Institute for Advanced Study at The Hong Kong
University of Science and Technology.  C.J Xia is supported by the
National Natural Science Foundation of
China (Grant No. 12275234 and No. 12342027) and the National SKA Program of China (No. 2020SKA0120300). R.-X Xu is supported by the National SKA Program of China (2020SKA0120100).

\end{acknowledgments}

\end{document}